\documentstyle[12pt]{article}

\title{\bf Signature change, vacuum condensation \\ and \\ cosmological constant }
\author{F. Darabi\thanks{e-mail:
f.darabi@azaruniv.edu} \\ {\small Department of Physics,
Azarbaijan University of Tarbiat Moallem , Tabriz, 53714-161
Iran.} }
\begin{document}
\maketitle
\begin{abstract}
We proposes an alternative model of duality symmetry, based on the
previously obtained divergence theory, including an scalar field, an
internal vector and a metric signature. At some small scale an
effective scalar field equation has appeared whose potential acts
like a Higgs one, where the metric signature plays the role of an
order parameter. Non-vanishing Vacuum condensation of this Higgs
field occurs once a signature change from Euclidean to Lorentzian is
formed. The mass scale of Higgs field excitations around this vacuum
may contribute, in the Lorentzian sector, to the cosmological
constant, in agreement with observations.

\end{abstract}
\newpage
\section{Introduction}

The initial idea of signature change was due to Hartle, Hawking
and Sakharov \cite{HHS}. This idea would make it possible to have
both Euclidean and Lorentzian metrics in the path integral
approach to quantum gravity. However, it was later shown that
signature change may happen, as well, in classical general
relativity \cite{CSC}.

Most of the works regarding the signature change dealt with
situations where the signature changing metric is defined {\it
apriori} on the manifold and one looks for the effects of the
assumed signature change on the Einstein equations or propagation
of particles in such a manifold. However, there are some other
viewpoints in which the signature generation of the space-time is
studied and considered to be a dynamical phenomenon at quantum
gravity regime. This is because a quantum formulation of
gravitation should accommodate geometries with degenerate metrics
and nontrivial topologies. However, the quantum nature of gravity
is quite different from the one of quantum field theory.
Therefore, in these viewpoints the question is how best to model
such a dynamical effect in the language of quantum field theory.

Percacci has offered a formalism in which one can dissociate the
conventional geometrical interrelations between the metric tensor
components and a field of co-frames, and work in close analogy
with the Higgs model in non-Abelian gauge theories \cite{Pre}.
Classical geometry is then regarded as an interpretation of
certain expectation values which minimize an effective action.
Greensite, on the other hand, has developed this idea further by
assuming that a particular pattern of signature arises dynamically
as a result of dynamical phase field which interpolates between
signatures \cite{Gre}. He has argued that, at least for the free
scalar field theory interacting with such a dynamical phase field,
the Lorentzian signature of a four-dimensional manifold can be
predicted as a ground state expectation value. Odintsov {\it et
al} have obtained the effective potential for the dynamical phase
field ( dynamical signature ) induced by the quantum effects of
massive fields on a topologically non-trivial $D$-dimensional
background, and shown that the Lorentzian signature is preferred
in both $D=6$ and $D=4$ \cite{Odin}.

From a different viewpoint, it was shown that the signature change
phenomena can be studied in a variant of the {\it divergence
theory}, from the viewpoint of a {\it time asymmetric law} in
vacuum which breaks an specific duality symmetry \cite{Sal}.  It
was based on the realization that a duality breaking must be
connected with the emergence of a preferred arrow of time in
vacuum, and this provides an indication that the time asymmetric
law may act to produce the condensation of vacuum.

In the present paper, following the above idea, we propose a new
duality symmetry different from the one introduced in \cite{Sal}.
We remark that at small distances, an effective Higgs type
potential may arise and show that the duality breaking manifests
directly as the condensation of vacuum due to a signature change
from Euclidean to Lorentzian. This model as an alternative to
\cite{Sal} has no specific advantage, but may deserve to be
studied, as well, to show the potential of divergence theory to
accommodate two different models of duality symmetry. Both models
interrelate the signature change problem to the vacuum
condensation of a scalar field, however, in two different ways of
duality breaking. Unlike \cite{Sal}, here, we introduce a duality
breaking as a {\it spontaneous symmetry breaking} and add a new
element as the cosmological constant originating from field
excitations around the vacuum and show that the cosmological
constant does not essentially change due to the signature change.

\section{The model}

It is well known that an exact Lorentz invariant vacuum of the
quantum field theory has, by definition, zero energy density. On the
other hand, the vacuum in quantum field theory is related to the
condensation of scalar fields leading to constant vacuum expectation
values for these fields. The appearance of these non-zero values and
the resulting mass scales contribute to the energy content of the
vacuum to increase its value from zero. The resulting non-vanishing
value accounts for a principle violation of Lorentz invariance. Such
a violation of Lorentz invariance may have origin in the unification
of gravity and quantum physics, and is expected to occur at
ultrashort distance regime described by an absolute length scale,
$l_0$. It is obvious that the existence of this universal length
scale is in sharp contrast with the universal requirement of Lorentz
invariance. In fact, in the Minkowski space-time, as a framework for
Lorentz invariance, there is no absolute line of demarcation between
large and small scales. However, if a positive definite measure of
distance is defined, then the Lorentz invariance will be broken
down. In so doing, we may follow the Blokhintsev point of view by
associating a time-like vector $N_{\mu}$ (the so called internal
vector) to the Minkowski space-time \cite{Blo}. In this way, one may
distinguish between small and large scales by taking the positive
definite interval
\begin{equation}
R^2=(2N_{\mu}N_{\nu} -\eta_{\mu \nu})x^{\mu}x^{\nu}, \label{1}
\end{equation}
where $\eta_{\mu \nu}=diag(1, -1, -1, -1)$ is the Minkowski metric
and $N_{\mu}=(1, 0, 0, 0)$ is a time-like vector. Given such a
positive definite metric as
\begin{equation}
\bar{\eta}_{\mu \nu}=2N_{\mu}N_{\nu}-\eta_{\mu \nu}, \label{2}
\end{equation}
it is then possible to determine the absolute size of a distance
by comparing $R$ with the universal length $l_0$.

One may then define the pair ($l_0, N_{\mu}$) whose physical
interpretation is as follows \cite{Blo, Sal}: $l_0$ is a
characteristic size in the ultrashort distance regime which acts
as a sort of universal length that determines a lower bound on any
scale of length probed in a realistic measurement. $N_{\mu}$
serves as a four-velocity of a preferred inertial observer in
vacuum by which the concept of the universal length $l_0$ makes
sense. Therefore, the pair ($l_0, N_{\mu}$) arises as the physical
feature of a Lorentz non-invariant vacuum.

We now introduce a divergence theory developed in Ref.\cite{Sal}.
This theory begins with a current
\begin{equation}
J_{\mu}=-\frac{1}{2}\phi
\stackrel{\leftrightarrow}{\partial_{\mu}}\phi^{-1}, \label{3}
\end{equation}
for which we obtain
\begin{equation}
\partial_{\mu}J^{\mu}=\phi^{-1}[\Box \phi
-\phi^{-1}\partial_{\mu}\phi \partial^{\mu}\phi], \label{4}
\end{equation}
and
\begin{equation}
J_{\mu}J^{\mu}=\phi^{-2}\partial_{\mu}\phi \partial^{\mu}\phi,
\label{5}
\end{equation}
where $\phi$ is a real scalar field. Combining these relations
leads to
\begin{equation}
\Box \phi + \Gamma\{ \phi \} \phi=0, \label{6}
\end{equation}
where $\Gamma\{ \phi \}$ is called the {\it dynamical mass term}
\begin{equation}
\Gamma\{ \phi \}=-J_{\mu}J^{\mu}-\partial_{\mu}J^{\mu}. \label{7}
\end{equation}
It is important to note that Eq.(\ref{6}) is a formal consequence
of the definition (\ref{3}), in the form of an identity, and is
not a dynamical equation for $\phi$. However, a large class of
dynamical theories may be obtained in the form of a {\it
divergence theory} by taking various currents $J^{\mu}$ in the
dynamical mass term. For example, a simple divergence theory is
developed by assuming
\begin{equation}
\partial_{\mu}J^{\mu}=0, \label{8}
\end{equation}
which leads, through the field redefinition $\sigma=\ln \phi$, to
\begin{equation}
\Box \sigma=0. \label{9}
\end{equation}
The dynamical mass term vanishes for this divergence theory and
the field $\sigma$ becomes massless. One may allow for a dynamical
coupling of $\phi$ with the internal vector $N_{\mu}$ by taking a
more complicate dynamical mass term. This is done by a divergence
equation of the type
\begin{equation}
\partial_{\mu}J^{\mu}=N_{\mu}N_{\nu}J^{\mu}J^{\nu}+g{l_0}N_{\mu}N_{\nu}N_{\sigma}J^{\mu}J^{\nu}J^{\sigma},\label{10}
\end{equation}
which leads to the field equation
\begin{equation}
\Box \phi
-(J_{\mu}J^{\mu}+N_{\mu}N_{\nu}J^{\mu}J^{\nu}+g{l_0}N_{\mu}N_{\nu}N_{\sigma}J^{\mu}J^{\nu}J^{\sigma})
\phi=0,\label{11}
\end{equation}
where $g$ stands for the metric signature which is positive for
Euclidean and negative for Lorentzian domains\footnote{The excess of
plus signs over minus signs is called the signature.}. Now, the
basic point is that the source of the divergence is invariant under
the following {\it dual} transformations\footnote{The source
(\ref{10}) and the dual transformations (\ref{12}) are modifications
of those introduced in Ref.\cite{Sal}. }
\begin{equation}
\phi \rightarrow -\phi \:\:\:,\:\:\:N_{\mu}\rightarrow -N_{\mu}
\:\:\:,\:\:\: g\rightarrow -g,\label{12}
\end{equation}
where the latter transformation accounts for a signature change
from Lorentzian to Euclidean
$$
\eta_{\mu \nu} \rightarrow \bar{\eta}_{\mu \nu}=
2N_{\mu}N_{\nu}-\eta_{\mu \nu},
$$
or vice versa. We note that the identity (\ref{6}) holds no matter
what signature is used in its derivation. Therefore, the field
equation (\ref{11}) holds for both Lorentzian and Euclidean
signatures. One may then assume a dynamical symmetry between the
dual configurations
\begin{equation}
(\phi, N_{\mu}, g)\Longleftrightarrow(-\phi, -N_{\mu},
-g),\label{13}
\end{equation}
so that the field equation (\ref{11}) makes no distinction between
these dual configurations. In fact, both signatures are related by
the dynamical symmetry of the field equation under the dual
transformations (\ref{12}). It is important to note that the
emergence of this duality in the field equation (\ref{11}) is
considered as reflecting the essential feature of the broken phase
of Lorentz invariance at small distance $\sim l_0$.

\section{Symmetry breaking}

In physics there are many models in which there are some
symmetries at the dynamical level leading to equivalent vacua in
the theory. No distinction between these equivalent vacua is
possible unless we require some symmetry breaking conditions to be
imposed on the physically realizable configuration of vacuum.
Usually, It is interesting to realize a symmetry breaking directly
by means of boundary conditions. But other approaches are also
possible and deserve to be studied, as well. For example, in
\cite{Sal} the authors try to establish a duality symmetry between
two configurations
\begin{equation}
(\phi, N_{\mu}, \eta_{\mu \nu})\Longleftrightarrow(\phi^{-1},
-N_{\mu}, \bar{\eta}_{\mu \nu}),\label{14}
\end{equation}
where $\eta_{\mu \nu}$ and $\bar{\eta}_{\mu \nu}$ stand for
Lorentzian and Euclidean metrics, respectively. They introduce
duality breaking by resorting to a {\it time asymmetric law} to be
imposed on a specific source of the divergence. The time
asymmetric law is then related to the vacuum condensation of the
quantum scalar field. The present model, although follows the same
purpose to take into account the signature change problem, but
takes different source of divergence (\ref{10}) and different
equivalent vacuum configurations (\ref{13}). Therefore, a
different duality breaking condition is required to realize the
physical vacuum configurations.

One alternative aspect of duality breaking seems to be a preferred
background value $\bar{\phi}$ and correspondingly a preferred
background direction of $N_{\mu}$, as average values taken over
large distances, which may be interrelated via a relation of the
type \cite{Sal}
\begin{equation}
\bar{J}_{\mu}=\frac{\lambda}{l_0}N_{\mu}, \label{15}
\end{equation}
where $\lambda$ measures the ratio of the universal length and the
spatial size of the universe as
\begin{equation}
\lambda=\frac{l_0}{R}.\label{15'}
\end{equation}
In this way, the duality breaking and correspondingly the preference
of a signature is considered to be a cosmological effect. If,
however, one is interested in realizing a preferred signature, it is
not unreasonable to expect that such a duality breaking should have
its origin at small distance regime rather than large cosmological
distances. This is because, the metric is, in principle, defined to
produce the most small distance appropriate for a realistic
measurement process. Therefore, it is more natural to think about
the origin and preference of Lorentzian over Euclidean metric as a
small distance effect. In this regard, unlike Ref.\cite{Sal}, we
consider Eq.(\ref{15}) not as a duality breaking but as a trick to
effectively linearize the source of divergence equation (\ref{10})
at ultrashort distance regime. This linearization is necessary in
order to compute the effective form of the dynamical mass term in
(\ref{11}) by obtaining an approximate solution for $J_{\mu}$.
Having this in mind, we can linearize the quadratic term in
$J_{\mu}$ in the source to find the approximations
\begin{equation}
\partial_{\mu}J^{\mu}=N_{\mu}N_{\nu}\bar{J}^{\mu}J^{\nu}+g{l_0}N_{\mu}N_{\nu}N_{\sigma}\bar{J}^{\mu}J^{\nu}J^{\sigma},
\label{16}
\end{equation}
and
\begin{equation}
\Box \phi
-(\bar{J}_{\mu}J^{\mu}+N_{\mu}N_{\nu}\bar{J}^{\mu}J^{\nu}+g{l_0}N_{\mu}N_{\nu}N_{\sigma}\bar{J}^{\mu}J^{\nu}J^{\sigma})
\phi=0.\label{17}
\end{equation}
We now truncate the nonlinear term in $J_{\mu}$ from the source of
the divergence (\ref{16}) and use Eq.(\ref{15}) to obtain
\begin{equation}
\partial_{\mu}J^{\mu}\simeq \frac{\lambda}{l_0}N_{\mu}J^{\mu}.
\label{18}
\end{equation}
Using Eq.(\ref{3}), we can write the divergence equation
(\ref{10}) in terms of $\phi$ as
\begin{equation}
\partial_{\mu}J^{\mu}\simeq \frac{\lambda}{l_0}N_{\mu}\frac{\partial^{\mu}\phi}{\phi}.\label{19}
\end{equation}
If we linearize this equation by inserting the average background
value of $\phi$ in the dominator we find an approximate solution
for $J^{\mu}$ as
\begin{equation}
J_{\mu}\simeq \frac{\lambda}{l_0}N_{\mu}\frac{\phi}{\bar{\phi}}.
\label{20}
\end{equation}
By using this solution for $J^{\mu}$ in (\ref{11}) we arrive at
the following field equation
\begin{equation}
\Box \phi -(2\frac{\lambda^2}{l_0^2}\frac{\phi}{\bar{\phi}}+g
\lambda \frac{\lambda^2}{l_0^2}\frac{\phi^2}{\bar{\phi}^2})
\phi=0.\label{21}
\end{equation}
If we replace ${\phi}$ by $\bar{\phi}$ in the linear term, we can
get a more effective form of this equation as
\begin{equation}
\Box \phi -(2\frac{\lambda^2}{l_0^2}+g \lambda
\frac{\lambda^2}{l_0^2}\frac{\phi^2}{\bar{\phi}^2})
\phi=0.\label{22}
\end{equation}
This equation can be derived from the following effective
potential
\begin{equation}
V(\phi)=-\frac{\lambda^2}{l_0^2}\phi^2-\frac{1}{4}g \lambda
\frac{\lambda^2}{l_0^2}\frac{\phi^4}{\bar{\phi}^2}. \label{23}
\end{equation}
This potential, has one minimum at $\phi_0=0$ and two degenerate
minima at $\phi_0=\pm \sqrt{\frac{2}{-g\lambda}}\bar{\phi}$
provided $g>0$ and $g<0$, respectively. This means, as far as the
signature of the metric is Euclidean, no duality breaking happens
to distinguish between dual configurations $\phi$ and $-\phi$.
When, however, a signature change happens from Euclidean to
Lorentzian the scalar field then condensates in one of the two
vacua $\phi_0=\pm \sqrt{\frac{2}{-g\lambda}}\bar{\phi}$ and one
configuration $(\phi, N_{\mu}, \eta_{\mu \nu})$ or $(-\phi,
-N_{\mu}, \bar{\eta}_{\mu \nu})$ is singled out,
permanently\footnote{This is what we meant by the duality breaking
at small distance regime. In fact, the smallness of $\lambda$
leads the two degenerate vacua to be too far apart from each
other, so the possibility of tunneling from one vacuum to the
other one almost vanishes.}. Therefore, the condensation of $\phi$
corresponds to a signature change from Euclidean to Lorentzian.

The ground state value $\phi_0$ of $\phi$ is obtained by
minimizing $V(\phi)$ which leads to the condition
\begin{equation}
\phi_0^2=\frac{2}{-g\lambda}\bar{\phi}^2.\label{24}
\end{equation}
If the potential $V(\phi)$ is expanded around $\phi_0$ we obtain (
neglecting constant terms )
\begin{equation}
V(\phi)= 4\left(\frac{\lambda}{l_0}\right)^2(\phi-\phi_0)^2+{\it O
}(\phi-\phi_0)^3, \label{25}
\end{equation}
from which one infers that physical excitations of $\phi$ around
$\phi_0$ have a preferred mass scale $m \sim \frac{\lambda}{l_0}$.
According to the quantum field theory considerations, any massive
particle excitation around a given scalar field vacuum can
contribute to the total value of the cosmological constant. But this
leads to the well-known cosmological constant problem. A solution of
this problem is to reduce the issue of the cosmological constant to
a picture in which the consistent contribution to the total value of
that constant comes from a {\it preferred mass scale} of the vacuum.
In this regard, the preferred mass scale $m \sim
\frac{\lambda}{l_0}$ may contribute to the cosmological constant, in
the Lorentzian sector, as $\Lambda \sim (\frac{\lambda}{l_0})^2 \sim
\frac{1}{R^2}$, where use has been made of Eq.(\ref{15'}). This is
in agreement with the present observational bound for the
cosmological constant as a remarkable consequence of the well-known
empirical fact that the present universe has just the characteristic
size $R \sim 10^{29}cm$. Note that according to Eq.(\ref{15'}) this
agreement with the observational bound for $\Lambda$ is obtained
merely by the ratio $\lambda/l_0$, and is independent of the
specific numerical values of $\lambda$ or $l_0$. The understanding
of the relation of $l_0$ to the Planck length is an elusive task of
quantum gravity. However, one must take a physically reasonable
estimation for the universal length. For example, $l_0$ may act as
the length scale that measures the size of the regime on which a
significant nonlinear self coupling like the Higgs potential
$V(\phi)$ can occur. This may constrain $l_0$ to be smaller than
electroweak or even supersymmetry scale.

It is worth noting that the vacuum energy as the minimum of
potential vanishes for the Euclidean metric, but the field
excitations around this minimum have almost the same mass scale as
that of Lorentzian sector. Therefore, they may contribute to the
cosmological constant in the Euclidean sector in the same manner as
in the Lorentzian sector. In other words, the cosmological constant
is almost invariant under the dual transformations (\ref{13}).

\section{Conclusion}

In this paper, we have studied a model of scalar field coupled to an
internal vector, in the presence of a universal length, having a
special duality symmetry in which the signature change appears as a
natural symmetry. When this dynamical symmetry is combined with a
duality breaking, a fixed background metric signature is singled
out. Contrary to similar models already introduced \cite{Sal} in
which the duality breaking arises due to large cosmological scale,
we have argued that this duality breaking may be established, as
well, at small distance regime by spontaneous symmetry breaking in
an effective potential which has been derived in the study of a
principle of duality invariance of the dynamical mass term of $\phi$
at a universal length in the small distance regime. It is shown that
the duality breaking as the condensation of the scalar field occurs
once a signature change from Euclidean to Lorentzian is happened at
small distance regime. This signature change leads to the emergence
of a preferred mass scale $\sim \frac{1}{R^2}$ of the vacuum which
contributes to the cosmological constant, in the Lorentzian sector.
The same mass scale is also appeared in the Euclidean sector which
leads to the same cosmological constant.

In the future, we hope to study and report more on the possible
deep relation between the signature change and spontaneous
symmetry breaking scenarios, since both of them are seriously
assumed to be happened at early universe.

\section*{Acknowledgment} This work has been supported
by the Research Department of Azarbaijan University of Tarbiat
Moallem, Tabriz, Iran.

\end{document}